\documentclass{article}%
\usepackage{amsmath}%
\usepackage{amsfonts}%
\usepackage{amssymb}%
\usepackage{graphicx}
\providecommand{\U}[1]{\protect\rule{.1in}{.1in}}

\begin{document}

\title{Massless two-loop ``master'' and \\ three-loop two point function in NDIM}
\author{A. T. Suzuki$^{1,a}$ \\
\\$^{1}$Instituto de F\'{\i}sica Te\'{o}rica Univ. Estadual Paulista\\Rua Dr. Bento Teobaldo Ferraz, 271 \\ 01140-070 - S\~{a}o Paulo, SP -- Brazil}
\maketitle

\begin{abstract}
NDIM (Negative Dimensional Integration Method) is a technique for evaluating Feynman integrals based on the concept of analytic continuation. The method has been successfully applied to many diagrams in covariant and noncovariant gauge field interactions and has shown its utility as a powerful technique to handle Feynman loop integrations in quantum field theories. In principle NDIM can handle any loop calculation; however, in practical terms, the resulting multiseries with several variables in general cannot be summed up conveniently and its analytic properties are generally unknown. 

The alternative then is to use order by order (loop by loop) integration in which the first integral is of the triangle diagram type. However, the na\"{i}ve momentum integration of this leads to wrong results. Here we use the shortened version for the triangle in NDIM that is suitable for a loop by loop calculation and show that it leads (after appropriate analytic continuation to positive dimension) to agreement with the known result for the two-loop master diagram. From it, a three-loop is then calculated and shown again its consistency with the already published result for such a diagram.

\vspace{.3cm}


\end{abstract}

\vspace{.3cm}

$^{a}${\footnotesize E-mail: suzuki@ift.unesp.br}

\section{Introduction}

Perturbative calculations play an important role in field theoretical approach to understanding particle interactions. Several techniques have been developed to tackle the ever increasing complexities for the task of evaluating multiloop Feynman diagrams, mostly in the context of  dimensional regularization \cite{dreg1} or analytic regularization \cite{cicuta} --- and among them we can mention the powerful Mellin-Barnes contour
integration \cite{mb1, mb2, mb3}, the method of Gegenbauer polynomials \cite{gegenbauer}, the differential equations technique \cite{remiddi} and others \cite{laporta, hathrell}. The NDIM developed by Halliday and Ricotta \cite{halliday} has shown itself as a reliable one when applied to the
calculation of diagrams of one- \cite{glover, revisited, esdras}, two- \cite{2loops} and multi-loops \cite{ivan, achievements}, with scalar and tensorial structures and in noncovariant gauges \cite{tensor-gauge}. One of the advantages of NDIM is that it allows us to avoid the often cumbersome parametric integrals, transfering the problem into easier solving systems of linear equations instead. Another advantage of NDIM is that the exponents of propagators are taken to be arbitrary integers, so that one can solve the general case for each type of graph. A severe drawback of this method is that as the number of loops increases, the number of systems of linear equations that must be solved grows to staggering heights. One would like then to work out higher loops via loop-by-loop calculation with inserted simpler results.   

The analytic result for the one-loop massless triangle Feynman diagram has been evaluated long ago by Boos and Davydychev \cite{boos} and since then reproduced in many different contexts, e.g., \cite{mb1, glover, esdras}. It is written in terms of a linear combination of four Appel \cite{Appel} hypergeometric functions of two variables $F_{4}(\alpha, \beta; \gamma, \gamma\,'\:|\:x,\,y)$, with $x=k^{2}/p^{2}$ and $y=(k-p)^{2}/p^{2}$, where $k$, $p$ and $k-p$ label the three external momenta that flow along the triangle's three external legs. This well-known result for the off-shell triangle, however, {\it is not valid} for every momentum; those momenta must be such that $|x|<1$, $|y|<1$ and $|\sqrt{x}|+|\sqrt{y}|<1$. In other words, the series is defined inside some region of convergence and for this reason the well-known result of Boos and Davydychev \cite{esdras, boos} can not be used in a loop by loop calculation \cite{jpsj}. Because a loop integration implies that the integrated momentum runs from minus infinity to plus infinity, we can easily see why the linear combination of four $F_4$'s with same variables will run into trouble within a loop by loop calculation. To solve this difficulty one needs to use the correct analytic expression for the triangle diagram which allows further integration on the momentum variables appearing within the Appel's functions. Fortunately, such suitable and shortened version for the triangle diagram result can be constructed, which is written in terms of only three Appel's hypergeometric functions $F_{4}$. This shortened and simplified form that is adequate for further momentum integration is obtained using the analytic continuation properties obeyed by the Appel's functions that preserve momentum conservation in the three legs of the triangle diagram \cite{canadian}.

The paper is outlined as follows: First we translate the $D$-dimensional Feynman integral for the master diagram of Figure \ref{Figure 1:} in the language of NDIM, then we calculate it using the negative dimensions technique. Once the NDIM result is obtained, we perform an appropriate analytic continuation to positive $D$-dimensionality to get our desired result. 

\section{The two-loop ``master'' self-energy integral}

\begin{figure}[h]
\centering
\includegraphics[height=3cm,width=7cm]{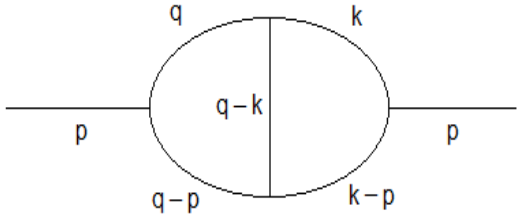}\caption{Two-loop two-point self-energy diagram}
\label{Figure 1:}
\end{figure}

Let us consider the two-loop ``master'' diagram as shown in Figure \ref{Figure 1:}. 
The Feynman integral associated to such a diagram is
\begin{equation}
\label{master}
I_{\rm master}=\int\int\frac{d^{D}k\;d^{D}q}{k^{2}(k-p)^{2}(k-q)^{2}
q^{2}(q-p)^{2}},
\end{equation}
where $p$ is the external momentum. 

In the NDIM we could, at least in principle, tackle this double integral in Eq. (\ref{master}) at once. However, as it has been already mentioned, this leads to the computation of more than eight thousand systems of coupled linear equations whose solutions are expressed as multiple series of hypergeometric-type with unknown analytic properties, so that we have no idea as to how to construct distinct complete sets of linearly independent solutions from among all those eight thousand plus results. 

Thus, the alternative is to evaluate it stepwise, order by order, performing one momentum integral at a time. To do this, we immediately note that there makes no difference at all whether we integrate first in $q$ or in $k$ since in both cases we have to integrate a one-loop triangle type integral first. We could write the above integral as, for example, 
\begin{equation}
I_{\rm master}=\int\frac{d^{D}q}{q^{2}(q-p)^{2}}\int\frac{d^{D}k}
{k^{2}(k-p)^{2}(k-q)^{2}},\label{1st_integral}
\end{equation}
and we readily recognize the integral in momentum $k$ as an off-shell triangle one, which has a well-known result \cite{boos} that can be written in terms of a linear combination of four Appel hypergeometric functions of two variables $F_{4}(\alpha,\, \beta;\, \gamma, \, \gamma\,'\:|\:x,\,y)$ with the two momentum dependent variables given, for example, as $x=(q-p)^{2}/p^{2}$ and $y=q^{2}/p^{2}$. Using now the series representation for the Appel functions of these specific two variables, we can therefore see that the remaining $q$-integral is of a self-energy type with shifted exponents for the propagators,
\begin{equation}
I_{\rm master}=\Gamma\int\frac{d^{D}q}{(q^{2})^{1+\mu}[(q-p)^{2}]^{1+\nu}
},\label{2nd_integral}
\end{equation}
where $\Gamma$ is a factor which depends on $p^{2}$, the dimension $D$ as well as the exponents of propagators. The shifting exponents $\mu$ and $\nu$ also depend on the dimension $D$ and exponents of propagators in the former integration, as well as on the double sum indices, say $a,b$, of the $F_{4}$ series. However, straightforward application of this does not yield the correct result. Here comes an important point: to carry out the second integral Eq.  (\ref{2nd_integral}) one has to perform the integral over the whole space, for this reason the result of the former one must hold on the whole range of momentum $q$. The well-known result of the off-shell triangle, written as a sum of four Appel's hypergeometric functions $F_{4}(...|x,y)$, {\it is not valid} for all momentum range; these momenta must be such that $|x|<1$, $|y|<1$ and $|\sqrt{x}|+|\sqrt{y}|<1$. In other words, the series is defined inside some region of convergence and for this reason the well-known result of Boos and Davydychev \cite{boos,esdras} can not be used in (\ref{1st_integral}) for the $k$ integration.

As demonstrated in \cite{canadian}, the result for the triangle diagram that should be plugged into Eq. (\ref{1st_integral}) has only three $F_4$'s. This choice guarantees that the $q$ integration may be performed for the whole interval $(-\infty < q <+\infty)$. 

\section{The two-loop ``master'' self-energy integral in NDIM}

The NDIM is characterized among other things, by two features: One is the generalized exponents for the propagators, say $(g, h, i, j, \cdots) \in \mathbb{N} $, and the other is the polynomial nature of the integrands that represent the propagators. Thus, in the spirit of the NDIM technique for performing Feynman integrals \cite{halliday, 2loops}, we introduce the NDIM counterpart of (\ref{1st_integral}), namely, 
\begin{equation}
I_{\rm master}^{\star(\rm NDIM)} \equiv \int d^{D}q\:(q^2)^{g}[(q-p)^2]^{h}\,\int d^{D}k
\:(k^2)^{i}\,[(k-p)^2]^{j}\,[(k-q)^2]^{l}.\label{1st_integralndim}
\end{equation}

Let us first concentrate our attention in the triangle part:
\begin{equation}
\label{Delta}
I^{\star(\rm {NDIM})}_{\Delta} \equiv\int {d^Dk}{(k^2)^i\,[(k-p)^2]^j\,[(k-q)^2]^l}.
\end{equation}

Since there is a recurrent appearance of a certain expression involving the exponents of propagators and dimension $D$, we introduce for convenience, the following definition, $\sigma \equiv i+j+l+D/2$,  and also define $r=q-p$. 

The standard solution for Eq. (\ref{Delta}) is a sum of four terms \cite{boos,esdras}:
\begin{eqnarray}
\label{triangle4}
I^{\star(\rm {NDIM})}_{\Delta}  & = & (-\pi)^{D/2}\sum_{n=1}^{4}\Lambda_n^4\,F_4\left(\alpha_n,\,\beta_n;\,\gamma_n,\,\gamma\,'_n\:|\: r^2/p^{2},\,q^2/p^{2}\right)
\end{eqnarray}
where the four coefficients $\Lambda_n^4$ are written in terms of Pochhammer's symbols $(a)_b = \Gamma(a+b)/\Gamma(b)$ of the several exponents of propagators and dimension $D$:
\begin{eqnarray}
\Lambda_1^4 & = & (p^2)^\sigma \frac{(1+\sigma)_{-2\sigma-D/2}}{(1+i)_{-\sigma}(1+j)_{-\sigma}}, \label{lambda1} \\
\Lambda_2^4 & = & (q^2)^\sigma\left(-\frac{p^2}{q^2}\right)^j\frac{(1+\sigma)_{-2\sigma-D/2}}{(1+i)_{-\sigma}(1+l)_{-\sigma}}\frac{(-\sigma)_j}{(1+l-\sigma)_j}, \label{lambda2}\\
\Lambda_3^4 & = & (r^2)^\sigma \left(-\frac{p^2}{r^2}\right)^i\frac{(1+\sigma)_{-2\sigma-D/2}}{(1+j)_{-\sigma}(1+l)_{-\sigma}}\frac{(-\sigma)_i}{(1+l-\sigma)_i}, \label{lambda3}\\
\Lambda_4^4 & = & \left( \frac{q^2 r^2}{p^2}\right)^\sigma \left(\frac{p^2}{r^2}\right)^i \left(\frac{p^2}{q^2}\right)^j\frac{(1-l-D/2)_{2l+D/2}}{(1+i)_{l+D/2}(1+j)_{l+D/2}}, \label{lambda4}
\end{eqnarray}
and the various parameters of the hypergeometric functions are listed in the table below:
\begin{table}[htbp]
\begin{center}
\begin{tabular}{|c||c|c||c|c|} \hline
n & $\alpha_n$ & $\beta_n$ & $\gamma_n$ & $\gamma^{\,\prime}_n$ \\
\hline \hline
$1$ & $-l$ & $-\sigma$ & $1+i-\sigma$ & $1+j-\sigma$ \\
\hline
$2$ & $-j$ & $-j-l+\sigma$ & $1+i-\sigma$ & $1-j+\sigma$ \\
\hline
$3$ & $-i$ & $-i-l+\sigma$ & $1-i+\sigma$ & $1+j-\sigma$ \\
\hline
$4$ & $l+D/2$ & $\sigma+D/2$ & $1-i+\sigma$ & $1-j+\sigma$\\
\hline
\end{tabular}
\caption{Parameters of Appel's functions in Eq. (\ref{triangle4}).}
\end{center}
\end{table}

Therefore Eq. (\ref{master}) becomes a sum of four terms if we use the standard solution \cite{boos,esdras} for the triangle diagram part:
\begin{equation}
\label{master4}
I^{\star(\rm {NDIM})}_{\rm master} = (-\pi)^{D/2}\sum_{n=1}^{4}I_{\rm SE}^{\star n},
\end{equation}
which means that we have now four integrals $(n=1,2,3,4)$ to solve, namely,
\begin{eqnarray}
I_{\rm SE}^{\star n} & \equiv &  \int d^Dq (q^2)^g\,(r^2)^h \Lambda_n^4 F_4(\alpha_n,\,\beta_n;\,\gamma_n,\,\gamma\,'_n\:|\: r^2/p^{2},\,q^2/p^{2})\\
                     & = & \sum_{a,b=0}^{\infty} \frac{(\alpha_n)_{a+b}(\beta)_{a+b}}{(\gamma_n)_a(\gamma\,'_n)_{b}}\frac{(p^2)^{-a-b}}{a! b!}\int d^Dq \,\Lambda_n^4 (q^2)^{g+b}\,(r^2)^{h+a}
\end{eqnarray}
where in the second line above we used the series representation for the Appel hypergeometric function of two variables $F_4$. Looking at Eqs. (\ref{lambda1})-(\ref{lambda4}) we see that the $q$-integration that remains is now very easy to perform since it is of the one-loop self-energy type, and for each $n$ we will have different exponents for $q^2$ and $r^2$ in the integrand, making different compositions as they join with the exponents $g+b$ and $h+a$ already presente in $q^2$ and $r^2$ respectively. 

If we follow this vein of calculation, what will happen is that after the analytic continuation to the positive dimension $D$ and negative specific values of exponents $g= h= i=j=l=-1$ , the ensuing result for Eq. (\ref{1st_integralndim})  is not concordant with the one calculated in positive dimensional integration by Chetyrkin {\em et al} \cite{gegenbauer} back in 1980. The reason behind this can be easily understood in that all four Appel functions that appear in the result of the first triangle diagram integration have the same two ratios of momentum variables defining the same convergence factor for all of them \cite{Appel, jpsj}. Further $q$-integration will of course violate this region of convergence aforementioned. We need therefore a modified version for the triangle diagram integral which will allow us performing the necessary next step $q$-integration. In our previous work, we have demonstrated what this modified triangle diagram integral is that can be embedded into higher order loop momentum integration \cite{canadian}. As before mentioned, in this case of the two-loop master diagram, the higher order loop momentum that needs to be calculated is of the one-loop self-energy type of integral.

The correct analytic expression for the triangle diagram that should be used in (\ref{1st_integralndim}) is given in \cite{canadian}
\begin{eqnarray}
\label{triangle3}
I^{\star(\rm {NDIM})}_{\Delta} & = & (-\pi)^{D/2}\sum_{n=1}^{2}\Lambda_n\,F_4(\alpha_n,\,\beta_n;\,\gamma_n,\,\gamma\,'_n\:|\: x,\,y) \nonumber \\
                                                   & + & (-\pi)^{D/2}\Lambda_{3}\,F_4(\alpha_{3},\,\beta_{3}\,\gamma_{3},\,\gamma\,'_{3}\:|\: \tilde{x},\,\tilde{y}),
\end{eqnarray}
where the {\it three} coefficients $\Lambda_n$ are:
\begin{eqnarray}
\Lambda_1 & = & (p^2)^\sigma \frac{(1+\sigma)_{-2\sigma-D/2}}{(1+i)_{-\sigma}(1+j)_{-\sigma}} \,, \label{lambda1A}\\
\Lambda_2 & = & (q^2)^\sigma\left(-\frac{p^2}{q^2}\right)^j\frac{(1+\sigma)_{-2\sigma-D/2}}{(1+i)_{-\sigma}(1+l)_{-\sigma}}\frac{(-\sigma)_j}{(1+l-\sigma)_j} \,,\label{lambda2A}\\
\Lambda_3 & = & (r^2)^\sigma \left(-\frac{q^2}{r^2}\right)^i\frac{(1+\sigma)_{-2\sigma-D/2}}{(1+j)_{-\sigma}(1+l)_{-\sigma}}\frac{(-\sigma)_i}{(1+j-\sigma)_i} \,. \label{lambda3A}
\end{eqnarray}

A conspicuous feature of this result is that the third hypergeometric function is now given in terms of {\it different} variables, i.e., $r^2/q^2$ and $p^2/q^2$ instead of the two previous ones $r^2/p^2$ and $q^2/p^2$. This set of solutions that is linearly combined allows for the regions of convergence to smoothly continue from $-\infty < q < +\infty$ as required by the necessary $q$-integration \cite{jpsj}.

Since the triangle diagram is invariant under external momentum exchanges, first $(p,j) \leftrightarrow (q,l)$, and from this symmetrized result a further exchange $(q,j) \leftrightarrow (r,i)$, such symmetries tell us that there are two additional sets of possible coefficients, namely, 
\begin{eqnarray}
\Lambda_{1^\prime} & = & (q^2)^\sigma \frac{(1+\sigma)_{-2\sigma-D/2}}{(1+i)_{-\sigma}(1+l)_{-\sigma}} \,, \label{lambda4A}\\
\Lambda_{2^\prime} & = & (p^2)^\sigma\left(-\frac{q^2}{p^2}\right)^l\frac{(1+\sigma)_{-2\sigma-D/2}}{(1+i)_{-\sigma}(1+j)_{-\sigma}}\frac{(-\sigma)_l}{(1+j-\sigma)_l} \,,\label{lambda5A}\\
\Lambda_{3^\prime} & = & (r^2)^\sigma \left(-\frac{p^2}{r^2}\right)^i\frac{(1+\sigma)_{-2\sigma-D/2}}{(1+j)_{-\sigma}(1+l)_{-\sigma}}\frac{(-\sigma)_i}{(1+l-\sigma)_i} \,, \label{lambda6A}
\end{eqnarray}
and
\begin{eqnarray}
\Lambda_{1^{\prime\prime}} & = & (r^2)^\sigma \frac{(1+\sigma)_{-2\sigma-D/2}}{(1+j)_{-\sigma}(1+l)_{-\sigma}} \,, \label{lambda7A}\\
\Lambda_{2^{\prime\prime}} & = & (p^2)^\sigma\left(-\frac{r^2}{p^2}\right)^{l}\frac{(1+\sigma)_{-2\sigma-D/2}}{(1+i)_{-\sigma}(1+j)_{-\sigma}}\frac{(-\sigma)_l}{(1+i-\sigma)_l} \,,\label{lambda8A}\\
\Lambda_{3^{\prime\prime}} & = & (q^2)^\sigma \left(-\frac{p^2}{q^2}\right)^j\frac{(1+\sigma)_{-2\sigma-D/2}}{(1+i)_{-\sigma}(1+l)_{-\sigma}}\frac{(-\sigma)_j}{(1+l-\sigma)_j} \,. \label{lambda9A}
\end{eqnarray}

The corresponding various parameters and variables of the Appel's hypergeometric functions of two variables are listed in the Table \ref{table2}.
\begin{table}[htbp]
\begin{center}
\begin{tabular}{|c||c|c||c|c||c|c|} \hline
n & $\alpha_n$ & $\beta_n$ & $\gamma_n$ & $\gamma^{\,\prime}_n$ & $x (\tilde{x})$ & $y (\tilde{y})$ \\
\hline \hline
$1$ & $-l$ & $-\sigma$ & $1+i-\sigma$ & $1+j-\sigma $ & $x=\frac{r^2}{p^2}$ & $y=\frac{q^2}{p^2}$ \\
\hline
$2$ & $-j$ & $-j-l+\sigma$ & $1+i-\sigma$ & $1-j+\sigma $ & $ x=\frac{r^2}{p^2}$ & $y=\frac{q^2}{p^2}$ \\
\hline
$3$ & $-i$ & $-i-j+\sigma$ & $1-i+\sigma$ & $1+l-\sigma $ & $ \tilde{x}=\frac{r^2}{q^2}$ & $\tilde{y}=\frac{p^2}{q^2}$\\
\hline
\hline
\hline
$1^\prime$ & $-j$ & $-\sigma$ & $1+i-\sigma$ & $1+l-\sigma $ & $x=\frac{r^2}{q^2}$ & $y=\frac{p^2}{q^2}$ \\
\hline
$2^\prime$ & $-l$ & $-j-l+\sigma$ & $1+i-\sigma$ & $1-l+\sigma $ & $ x=\frac{r^2}{q^2}$ & $y=\frac{p^2}{q^2}$ \\
\hline
$3^\prime$ & $-i$ & $-i-l+\sigma$ & $1-i+\sigma$ & $1+j-\sigma $ & $ \tilde{x}=\frac{r^2}{p^2}$ & $\tilde{y}=\frac{q^2}{p^2}$\\
\hline
\hline
\hline
$1^{\prime\prime}$ & $-i$ & $-\sigma$ & $1+j-\sigma$ & $1+l-\sigma $ & $x=\frac{q^2}{r^2}$ & $y=\frac{p^2}{r^2}$ \\
\hline
$2^{\prime\prime}$ & $-l$ & $-i-l+\sigma$ & $1+j-\sigma$ & $1-l+\sigma $ & $ x=\frac{q^2}{r^2}$ & $y=\frac{p^2}{r^2}$ \\
\hline
$3^{\prime\prime}$ & $-j$ & $-j-l+\sigma$ & $1-j+\sigma$ & $1+i-\sigma $ & $ \tilde{x}=\frac{q^2}{p^2}$ & $\tilde{y}=\frac{r^2}{p^2}$\\
\hline
\hline
\end{tabular}
\caption{Parameters and variables of Appel's functions in Eq. (\ref{triangle3})} \label{table2}
\end{center}
\end{table}

In principle any one of these three sets (unprimed, primed or double primed) could be inserted into the remaining $q$-integral and the integration carried out. However, in order to compare our ensuing result for the two-loop ``master'' self-energy diagram with already known result, it so happens that the most convenient sets are the primed and/or the double primed ones. The reason why the unprimed set is not convenient is due to the fact that $\Lambda_3$ is proportional to both $q^2$ and $r^2$, which when inserted into the $q$-loop integration will lead to more complicate structure for the parameters in the function part of the answer. In our present case, this function part is expressed as a series of the form:
\begin{eqnarray}
I_{\rm master} \!\!\!\!& \propto & \!\!\!\!\sum_{m=0}^{\infty} \frac{(a)_m (b)_m (c)_m (d)_m}{(x)_m (y)_m (z)_m m!}{}_4F_3 \left(\!\!\!\!\!\!\!\left. \begin{array}{c} \begin{array}{cccc} a+m,& b+m,& e, & f  \end{array} \\
                                                            \begin{array}{ccc}  w,& x+m, & y+m  \end{array} \end{array}\!\!\!\!\!\right|1 \right)\!,
\end{eqnarray}
where parameters $(a,b,c,d,e, f; x,y,z, w)$ depend on the exponents of $q^2$ and $r^2$. When any one of the $(a,b,c,d)$ is zero, only the $m=0$ term in the sum survives and in this case ${}_4F_3(a,b,e,f;w,x,y|1)$ may coalesce into simpler forms ${}_2F_1(\alpha,\beta;\gamma|1)$ and thus be summed up using Gauss' summation formula for hypergeometric series after integration. This kind of sum simplification and coalescence of ${}_4F_3$ fails to occur in the case of $\Lambda_3$. Thus in the following we take the primed set of solutions for the triangle diagram to perform the calculations.

The $q$-momentum integral is of a one-loop self-energy type integral, so just for reference we quote here the general result valid for such an integral in NDIM:
\begin{eqnarray}
\label{se}
I_{\rm SE} & \equiv & \int d^Dq (q^2)^e [(q-p)^2]^f \nonumber \\
                & = & (-\pi)^{D/2}(p^2)^{\sigma_1} \frac{(1+\sigma_1)_{-2\sigma_1-D/2}}{(1+e)_{-\sigma_1}(1+f)_{-\sigma_1}}\,, \qquad \sigma_1 \equiv e + f + D/2.
\end{eqnarray}  

Application of this NDIM formula will produce again recurring expressions which we define conveniently using short hand notations, such that, for example, $\sigma^\prime \equiv g+h+D/2$ and $\Omega \equiv \sigma+\sigma^\prime = g+h+i+j+l+D$.  Thus, proper evaluation (details are left to Appendix) of those relevant three terms gives respectively:
\begin{eqnarray}
I_{\rm SE}^{\star 1} & = & (-\pi)^{D/2}(p^2)^{\Omega}\frac{(1+\sigma)_{-2\sigma-D/2}}{(1+i)_{-\sigma}(1+l)_{-\sigma}}\frac{(1+\Omega)_{-2\Omega - D/2}}{(1+g+\sigma)_{-\Omega} (1+h)_{-\Omega}} \nonumber \\
                                  && \times \sum_{m=0}^{\infty} \frac{(-j)_m(-\sigma)_m(1+h)_m(-g
+\sigma^\prime)_m}{(1+i-\sigma)_m(-g-\sigma)_m(1+h-\Omega)_m}\frac{1}{m!} \nonumber \\
&& \times {}_4F_3\left(\!\!\!\!\!\!\left. \begin{array}{c} \begin{array}{cccc} -j+m,& -\sigma+m,& -\Omega, & 1-\Omega-D/2  \end{array} \\
                                                            \begin{array}{ccc}  1+l-\sigma,& -g-\sigma+m, & 1+h-\Omega+m  \end{array} \end{array}\!\!\!\!\!\right|1 \right), \\
I_{\rm SE}^{\star 2} & = & (-\pi)^{D/2}(p^2)^{\Omega}\frac{(1+\sigma)_{-2\sigma-D/2}}{(1+i)_{-\sigma}(1+j)_{-\sigma}}\frac{(-1)^l(-\sigma)_l}{(1+j-\sigma)_l}\nonumber \\
&& \times \frac{(1+\sigma^\prime+l)_{-2\sigma^\prime-2l - D/2}}{(1+g+l)_{-\sigma^\prime-l} (1+h)_{-\sigma^\prime-l}} \nonumber \\
                                  && \times \sum_{m=0}^{\infty} \frac{(-l)_m(-j-l+\sigma)_m(1+h)_m(-g+\sigma^\prime)_m}{(1+i-\sigma)_m(1+h-\sigma^\prime-l)_m(-g-l)_m}\frac{1}{m!} \nonumber \\
&& \times {}_4F_3\left(\!\!\!\!\!\!\left. \begin{array}{c} \begin{array}{cccc} \!-l\!+\!m,& \!-j\!-\!l\!+\!\sigma\!+\!m,& \!1\!-\!\sigma^\prime\!-\!l-\!D/2, & -\sigma^\prime\!-\!l  \end{array} \\
                                                            \begin{array}{ccc}  1-l+\sigma,& 1+h-\sigma^\prime-l+m, & -g-l+m  \end{array} \end{array}\!\!\!\!\!\right|1 \right), \\
I_{\rm SE}^{\star 3} & = & (-\pi)^{D/2}(p^2)^{\Omega}\frac{(1+\sigma)_{-2\sigma-D/2}}{(1+j)_{-\sigma}(1+l)_{-\sigma}}\frac{(-1)^i(-\sigma)_i}{(1+l-\sigma)_{i}}\nonumber \\ 
&& \times \frac{(1+\Omega-i)_{-2\Omega+2i-D/2}}{(1+g)_{-\Omega+i}(1+h+\sigma-i)_{-\Omega+i}}  \nonumber \\
                                  && \times \sum_{m=0}^{\infty} \frac{(-j)_m(-j-l+\sigma)_m(1+g+\sigma-j)_m(-h+\Omega-j)_m}{(1-j+\sigma)_m(1+\Omega-j)_m(\Omega-j+D/2)_m}\frac{1}{m!} \nonumber \\
&& \times {}_4F_3\left(\!\!\!\!\!\!\!\left. \begin{array}{c} \begin{array}{cccc} -i\!+\!m,& \!\!-\!i-\!l\!+\!\sigma\!+\!m,& \!\!1\!+\!h\!+\!\sigma\!-\!i, &\!\! -g\!+\!\Omega\!-\!i  \end{array} \\
                                                            \begin{array}{ccc}  1+j-\sigma,& 1+\Omega-i+m, & \Omega-i+D/2+m  \end{array} \end{array}\!\!\!\!\!\right|1 \right).
\end{eqnarray}

We can now collect all the individual results and write our answer to Eq. (\ref{1st_integralndim}) as the linear combination
\begin{equation}
\label{result}
I_{\rm master}^{\star(\rm NDIM)}(g,h,i,j,l;D)=(-\pi)^D\left(p^2\right)^\Omega \left\{\mathbb{C}_1{\cal F}_1 +\mathbb{C}_2{\cal F}_2 +\mathbb{C}_3{\cal F}_3 \right\},
\end{equation}
where the three coefficients $\mathbb{C}_n$ and three  functions ${\cal F}_n$ are respectively given by:
\begin{eqnarray}
\mathbb{C}_1 & = & \frac{(1+\sigma)_{-2\sigma-D/2}}{(1+i)_{-\sigma}(1+l)_{-\sigma}}\frac{(1+\Omega)_{-2\Omega - D/2}}{(1+g+\sigma)_{-\Omega} (1+h)_{-\Omega}}, \label{c1} \\
\mathbb{C}_2 & = & \frac{(1+\sigma)_{-2\sigma-D/2}}{(1+i)_{-\sigma}(1+j)_{-\sigma}}\frac{(-1)^l(-\sigma)_l}{(1+j-\sigma)_l}\nonumber \\
&& \times \frac{(1+\sigma^\prime+l)_{-2\sigma^\prime-2l - D/2}}{(1+g+l)_{-\sigma^\prime-l} (1+h)_{-\sigma^\prime-l}}, \label{c2} \\
\mathbb{C}_3 & = & \frac{(1+\sigma)_{-2\sigma-D/2}}{(1+j)_{-\sigma}(1+l)_{-\sigma}}\frac{(-1)^i(-\sigma)_i}{(1+l-\sigma)_{i}}\nonumber \\ 
&& \times \frac{(1+\Omega-i)_{-2\Omega+2i-D/2}}{(1+g)_{-\Omega+i}(1+h+\sigma-i)_{-\Omega+i}}, \label{c3}
\end{eqnarray}
and 
\begin{eqnarray}
{\cal F}_1 & = & \sum_{m=0}^{\infty}\frac{(-j)_{m}(-\sigma)_m(1+h)_m(-g+\sigma^\prime)_m}{(1+i-\sigma)_{m}(1+h-\Omega)_{m}(-g-\sigma)_m}\frac{1}{m!} \nonumber \\
&& \times {}_4F_3\left(\!\!\!\!\!\!\!\left. \begin{array}{c} \begin{array}{cccc} -j+m,& -\sigma+m,& 1-\Omega-D/2, & -\Omega  \end{array} \\
                                                            \begin{array}{ccc}  1+l-\sigma,& 1+h-\Omega+m, & -g-\sigma+m  \end{array} \end{array}\!\!\!\!\!\!\right|1 \right) , \label{F1} \\
{\cal F}_2 & = & \sum_{m=0}^{\infty}\frac{(-l)_{m}(j-l+\sigma)_m(1+h)_m(-g+\sigma^\prime)_m}{(1+i-\sigma)_{m}(1+h-\sigma^\prime-l)_{m}(-g-l)_m}\frac{1}{m!}\nonumber \\
&& \times  {}_4F_3\left(\!\!\!\!\!\!\!\left. \begin{array}{c} \begin{array}{cccc} -l+m,& \!\!-\!j\!-l\!+\!\sigma\!+\!m,& \!1\!-\!\sigma^\prime\!-\!l-\!D/2, &\! \!-\sigma^\prime\!-\!l  \end{array} \\
                                                            \begin{array}{ccc}  1-l+\sigma,& 1+h-\sigma^\prime+m, & -g-l+m  \end{array} \end{array}\!\!\!\!\!\!\right|1 \right) ,  \label{F2} \\
{\cal F}_3  & = & \sum_{m=0}^{\infty}\frac{(-i)_{m}(-i-l+\sigma)_m(1+h+\sigma-i)_m(-g+\Omega-i)_m}{(1-i+\sigma)_{m}(1+\Omega-i)_{m}(\Omega-i+D/2)_m}\frac{1}{m!}\nonumber \\
&& \times {}_4F_3\left(\!\!\!\!\!\!\!\left. \begin{array}{c} \begin{array}{cccc} -i+m,& -i-l+\sigma+m,& 1+g, & -h+\sigma^\prime  \end{array} \\
                                                            \begin{array}{ccc}  1+j-\sigma,& 1+\Omega-i+m, & \Omega-i+D/2+m  \end{array} \end{array}\!\!\!\!\!\!\right|1 \right) .\label{F3}
\end{eqnarray}

Having obtained these results we now have to analytic continue to positive dimension and negative values of exponents $-g,-h,-i,-j,-l \in \mathbb{N}$. This is accomplished by operating on the Pochhammer's symbols present in the coefficients $\mathbb{C}_n$, which is typical of the NDIM technique.Then analyticly continuing Eq. (\ref{1st_integralndim}) we get from Eq. (\ref{result}):
\begin{eqnarray}
\label{ac}
I_{\rm master} & = & I_{\rm master}^{\star(\rm NDIM);AC} \nonumber \\
& = & (-\pi)^D\left(p^2\right)^\Omega \left\{\mathbb{C}_1^{\rm AC}{\cal F}_1 +\mathbb{C}_2^{\rm AC}{\cal F}_2 +\mathbb{C}_3^{\rm AC}{\cal F}_3 \right\}.
\end{eqnarray}

The final result for the two-loop ``master'' self-energy diagram with generalized exponents of propagators is therefore given by Eq. (\ref{ac}). In it the analytic continuation of the exponents to negative values, $-g, -h, -i,-j,-l$, in the Pochhammer's symbols are done using the well-known relation \cite{slater}:
\begin{equation}
(a)_n = \frac{(-1)^n}{(1-a)_{-n}}\,, \qquad a\in {\mathbb{Z}}.
\end{equation}

Using this analytic continuation relation for the various Pochhammer's factors in the coefficients given in Eqs. (\ref{c1})-(\ref{c3}) we get
\begin{eqnarray}
\mathbb{C}_1^{\rm AC}  & = & (-1)^D\frac{(-i)_{\sigma}(-j)_\sigma}{(-\sigma)_{2\sigma+D/2}}\frac{(-g)_{\sigma^\prime}(-h)_{\sigma^\prime}}{(-\sigma^\prime)_{2\sigma^\prime+D/2}}, \\
\mathbb{C}_2^{\rm AC} & = & (-1)^D\frac{(-i)_{\sigma}(-l)_{\sigma}}{(-\sigma)_{2\sigma+D/2}}\frac{(-g-\sigma)_{\Omega}(-h)_\Omega}{(-\Omega)_{2\Omega+D/2}}, \\
\mathbb{C}_3^{\rm AC} & = & (-1)^D\frac{(-i)_{\sigma}(-j)_{\sigma}}{(-\sigma)_{2\sigma
+D/2}}\frac{(-j+\sigma)_{-l}}{(l-\sigma)_{-l}}\frac{(l-\sigma)_{-j-l+\sigma}}{(j-\sigma)_{-j-l+\sigma}}\nonumber \\
       && \times \frac{(-g-\sigma+j)_{\Omega-j}\,(-h)_{\Omega-j}}{(-\Omega+j)_{2\Omega-2j+D/2}},
\end{eqnarray}

In order for us to check whether this result is consistent with known results previously obtained via other methods, it is necessary to particularize the result in Eq. (\ref{ac}) for the specific values $g=h=i=j=l=-1$, which is tantamount to evaluate the original Feynman integral for the two-loop master integral given in Eq. (\ref{master}). 

Moreover, as pointed out previously, since we are taking the special case $g = h = -1$, this implies that all sum terms where we meet $(1+h)_m=(0)_m$ vanishes except for $m=0$. Also $1+g$ as the numerator parameter in ${}_4F_3$ reduces the function to just a constant equal to 1.The other functions ${}_4F_3$ in Eqs. (\ref{F1}) and (\ref{F2}) will present the coincident numerator and denominator parameters so that they both coaslesce into two Gauss hypergeometric functions ${}_2F_1(a,b;c|1)$ with unity argument: 
\begin{eqnarray}
I_{\rm master} & = & \pi^D\left(p^2\right)^{D-5}\frac{\Gamma^2(1+\sigma)\Gamma(-\sigma)}{\Gamma(\sigma+D/2)}   \nonumber \\
&\times& \left[\frac{\Gamma(1+\sigma^\prime)\Gamma(1+\Omega)\Gamma(-\Omega)}{\Gamma(1-\sigma)\Gamma(\Omega+D/2)} {}_2F_1\left(\!\!\!\!\!\!\!\left. \begin{array}{c} \begin{array}{cc} 1,& 1-\Omega-D/2 \end{array} \\
                                                            \begin{array}{c}  1-\sigma  \end{array} \end{array}\!\!\!\!\!\!\right|1 \right) \right. \nonumber \\
&& 
- \frac{\Gamma(1+\sigma^\prime)\Gamma(\sigma^\prime)\Gamma(1-\sigma^\prime)}{\Gamma(\sigma^\prime-1+D/2)}{}_2F_1\left(\!\!\!\!\!\!\!\left. \begin{array}{c} \begin{array}{cc} 1,& 2-\sigma^\prime-D/2  \end{array} \\
                                                            \begin{array}{c}  2 \end{array} \end{array}\!\!\!\!\!\!\right|1 \right) \nonumber\\
&& \left. -\frac{\Gamma(2+\Omega)\Gamma(1+\sigma^\prime)\Gamma(-\Omega-1)}{\Gamma(-\sigma)\Gamma(\Omega+1+D/2)}{}_2F_1\left(\!\!\!\!\!\!\!\left. \begin{array}{c} \begin{array}{cc} 1,& 1+\sigma \end{array} \\
                                                            \begin{array}{c}  \Omega+1+D/2   \end{array} \end{array}\!\!\!\!\!\!\right|1 \right) \right].
\end{eqnarray}

The Gauss hypergeometric function of unity can be summed up using the Gauss summation formula \cite{slater}, 
\begin{equation}
{}_2F_1(a,b;c|1) = \frac{\Gamma(c)\Gamma(c-a-b)}{\Gamma(c-a)\Gamma(c-b)}.
\end{equation}

Using
\begin{eqnarray}
\sigma & = & D/2 - 3, \nonumber \\
\sigma^\prime & = & D/2 - 2, \\
\Omega & = & D - 5\,.\nonumber
\end{eqnarray}
we finally get
\begin{eqnarray}
I_{\rm master} & = & \pi^D\left(p^2\right)^{D-5}\frac{\Gamma^2(D/2-2)\Gamma(3-D/2)}{\Gamma(D-3)}   \nonumber \\
&\times& \left[\frac{\Gamma(D/2-1)\Gamma(D-4)\Gamma(5-D)\Gamma(D-3)}{\Gamma(3D/2-5)\Gamma(3-D/2)\Gamma(D-2)} \right. \nonumber \\
&& 
- \frac{\Gamma(D/2-1)\Gamma(D/2-2)\Gamma(3-D/2)}{\Gamma(D-2)} \nonumber\\
&& \left. -\frac{\Gamma(D-3)\Gamma(D/2-1)\Gamma(4-D)\Gamma(D-3)}{\Gamma(3-D/2)\Gamma(3D/2-5)\Gamma(D-2)} \right].
\end{eqnarray}

We may rewrite these three terms in a more compact form using the well known identity $n\Gamma(n)=\Gamma(n+1)$ and its variants. The second term within the square brackets may be written as
\begin{equation}
- \frac{\Gamma(D/2-1)\Gamma(D/2-2)\Gamma(3-D/2)}{\Gamma(D-2)} = - \frac{\Gamma^2(D/2-1)\Gamma(2-D/2)}{\Gamma(D-2)},
\end{equation}
while the first and third terms within square brackets may be written together as
\begin{equation}
-\frac{\Gamma^2(D-3)\Gamma(2-D/2)\Gamma(D/2-1)\Gamma(5-D)}{\Gamma^2(3-D/2)\Gamma(3D/2-5)\Gamma(D-2)}.
\end{equation}

Thus finally
\begin{eqnarray}
I_{\rm master} & = & \pi^D\left(p^2\right)^{D-5}\frac{\Gamma^2(D/2-2)\Gamma(2-D/2)\Gamma(D/2-1)}{\Gamma(D-2)}   \nonumber \\
&\times& \left[\frac{\Gamma(3-D/2)\Gamma(D/2-1)}{\Gamma(D-3)} - \frac{\Gamma(D-3)\Gamma(5-D)}{\Gamma(3-D/2)\Gamma(3D/2-5)}\right].
\end{eqnarray}
This is exactly the result obtained via Gegenbauer polynomials method by Chetyrkin's {\em et al} (cf. Eq. (2.14), p. 351 in the first reference in \cite{gegenbauer}).

\section{Three-loop two-point self energy in NDIM}

After obtaining the two-loop ``master'' self-energy diagram result, it is not difficult to get the three loop two-point funtion for the diagram depicted in Figure \ref{Figure2:}. We explicitly draw in it the momentum flow in each line and use the convention $s = q -k $, $R = Q+k-q=Q-s$ for convenience. 
\begin{figure}[h]
\centering
\includegraphics[height=3cm,width=7cm]{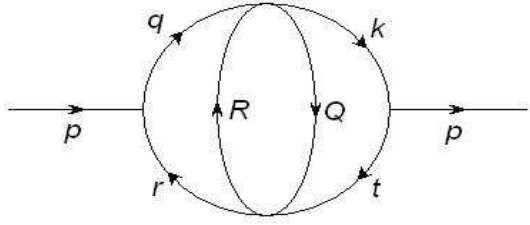}\caption{Three-loop two-point self-energy diagram}
\label{Figure 2:}
\end{figure}

Then the corresponding integral reads: 
\begin{equation}
\label{3loops}
I_{\rm 3 loops}^{\star (\rm NDIM)} = \int \int \int d^Dq\,d^Dk\,d^DQ (q^2)^g (r^2)^h (k^2)^i (t^2)^j(Q^2)^e (R^2)^f.
\end{equation} 

In the spirit of order-by-order integration, this can be written as
\begin{equation}
\label{3loops_dec}
I_{\rm 3 loops}^{\star (\rm NDIM)} = \int  d^Dq(q^2)^g (r^2)^h \int d^Dk(k^2)^i (t^2)^j\int d^DQ (Q^2)^e [(Q-s)^2]^f,
\end{equation} 
where in the first $Q$-integration we have explicited $R=Q-s$. This integral is a one-loop self-energy integral given in Eq. (\ref{se}) with $s$ in place of $p$ and $\sigma_1 = e+f+D/2$. Plugging this result into Eq. (\ref{3loops_dec}) we get
\begin{eqnarray}
\label{3loops_sigma1}
I_{\rm 3 loops}^{\star (\rm NDIM)} & = & (-\pi)^{D/2}\frac{(1+\sigma_1)_{-2\sigma_1-D/2}}{(1+e)_{-\sigma_1}(1+f)_{-\sigma_1}}\nonumber \\
                                                     & \times &  \!\!\! \int \!  d^Dq(q^2)^g [(q-p)^2]^h \int \! d^Dk(k^2)^i [(k-p)^2]^j [(k-q)^2]^{\sigma_1}.
\end{eqnarray} 

If we compare Eq. (\ref{3loops_sigma1}) with Eq. (\ref{1st_integralndim}) we note that the former has exactly the same structure of integrand as the latter one; the only difference being the exponent that appears in the $(q-k)^2$ factor --- instead of $l$ in the former, we have now $\sigma_1$ in the latter expression. Then we can immediately write
\begin{eqnarray}
I_{\rm 3 loops}^{\star (\rm NDIM)} & = & (-\pi)^{D/2}\frac{(1+\sigma_1)_{-2\sigma_1-D/2}}{(1+e)_{-\sigma_1}(1+f)_{-\sigma_1}}\nonumber \\
                                                     & \times &  I_{\rm master}^{\star (\rm NDIM)}(g,h,i,j,\sigma_1).
\end{eqnarray}

Upon analytic continuation to positive dimension and negative values of exponents, we get
\begin{eqnarray}
I_{\rm 3 loops} & = & \pi^{D/2}\frac{(-e)_{\sigma_1}(-f)_{\sigma_1}}{(-\sigma_1)_{2\sigma_1+D/2}}\times I_{\rm master}(g,h,i,j,\sigma_1).
\end{eqnarray}

It is now a simple matter of substituting the correct values of exponents for the case $e=f=g=h=i=j=-1$ and careffully manipulating the various gamma functions that appear to obtain the final result for the three loop integral
\begin{eqnarray}
\label{3loops_result}
I_{\rm 3 loops} & = & \int \int \int \frac{d^Dq\,d^Dk\,d^DQ}{ q^2 (q-p)^2 k^2 (k-p)^2 Q^2(Q-s)^2} .
\end{eqnarray}

The final result for Eq. (\ref{3loops_result}) is then
\begin{eqnarray}
I_{\rm 3 loops} & = & 2\pi^{3D/2}(p^2)^{3D/2-6}\frac{\Gamma^3(D/2\!-\!1)\Gamma(5\!-\!3D/2)}{(D\!-\!3)} \nonumber \\
& \times &\left\{ \cos(\pi D) \Gamma(2\!-\!D/2)\Gamma(3\!-\!D)\!-\!\frac{\Gamma(D/2\!-\!1)}{(3D/2-4)\Gamma(2D\!-\!5)} \right. \nonumber \\
&& \times \left. {}_3F_2\left(\!\!\!\!\!\!\!\left. \begin{array}{c} \begin{array}{ccc} 1,& D-2,& 3D/2-4  \end{array} \\
                                                            \begin{array}{cc}  2D-5,& 3D/2-3   \end{array} \end{array}\!\!\!\!\!\right|1 \right)\right\}.
\end{eqnarray} 

This is concordant with the result given in Hathrell \cite{hathrell} (see Eq. (8.13) on page 176).

\section{Conclusion}

In this work we have demonstrated that in a loop by loop calculation of higher order Feynman diagrams, the standard analytic solution for the one-loop triangle diagram expressed in terms of a linear combination of four Appel's hypergeometric functions of two variables cannot be used. The reason why such a solution for the triangle cannot be used can be understood considering that those variables defining the hypergeometric functions are restricted to convergence constraints, and loop integrations require momentum running from minus infinity to plus infinity. Also, our result hints that such an analytic expression for the triangle diagram also is not correct for further integration due to the fact that those variables are connected by a momentum conservation constraint, namely, $r = q-p$, and therefore, not all the four Appel's hypergeometric functions are linearly independent to each other. Therefore, on the grounds of mathematical argumentation concerning constraints to lower the number of independent functions as well as the physical argumentation connected to the domain of momentum integration and variables convergence region, the analytic function for the triangle diagram that allows for further momentum integration must be a linear combination of three independent Appel's function. Which three of these functions should be can only be determined invoking another physical input beyond momentum conservation \cite{canadian}. 

Using the modified triangle diagram integral expressed as a linear combination of {\it three} linearly independent solutions in terms of {\it three} Appel's functions of two variables, two of which have the same variables $x$ and $y$ and the third one having variables $x/y$ and $1/y$, \cite {canadian} we were able to calculate the two-loop ``master'' self-energy diagram using NDIM performing order-by-order calculation. Of course, our calculation shows that the same care must be taken for order-by-order calculation done in the usual positive dimensional calculations involving embedded triangle diagrams. 

Once the result for the two-loop ``master'' diagram is obtained, it is a matter of straighforward calculation to obtain the corresponding three-loop diagram as in Figure \ref{Figure 2:} since the Feynman integral associated to such diagram can be reduced to the two-loop case once a convenient one-loop self-energy diagram integral is performed. The only novelty is that this ensuing two-loop ``master'' integral now bears a shifted exponent in one of the integrand factors. The remaining of the calculation is just manipulation of gamma function factors using the property $n\Gamma(n) = \Gamma(n+1)$ and its related versions together with use of well-known properties of hypergeometric functions ${}_4F_3(\alpha_1,\alpha_2,\alpha_3,\alpha_4; \beta_1,\beta_2,\beta_3|1) $ and ${}_3F_2(\alpha_1,\alpha_2,\alpha_3; \beta_1,\beta_2|1)$ of unity argument.

\end{document}